\documentclass[acmsmall]{acmart}
\usepackage{lscape}
\usepackage{xcolor}
\usepackage{array}

\AtBeginDocument{%
  }

\setcopyright{cc}
\setcctype{by}
\acmJournal{PACMHCI}
\acmYear{2025}
\acmVolume{9}
\acmNumber{7}
\acmArticle{CSCW445}
\acmMonth{11}
\acmDOI{10.1145/3757626}

\urldef{\Wikipedia}\url{https://en.wikipedia.org/wiki/Wikipedia:Don%27t_cite_Wikipedia_on_Wikipedia}
\urldef{\Wikipediab}\url{https://en.wikipedia.org/wiki/Wikipedia:Categories,_lists,_and_navigation_templates#Navigation_templates}

\begin{document}

\title{Machines in the Margins: A Systematic Review of Automated Content Generation for Wikipedia}

\author{Neal Reeves}
\email{neal.t.reeves@kcl.ac.uk}
\orcid{0000-0002-1044-3943}
\affiliation{
    \institution{Department for Informatics, King's College London}
    \city{London}
    \country{United Kingdom}}
\author{Elena Simperl}
\orcid{0000-0003-1722-947X}
\affiliation{%
  \institution{Department for Informatics, King's College London}
  \city{London}
  \country{United Kingdom}
}
\affiliation{
    \institution{Institute for Advanced Study, Technical University of Munich}
    \city{Munich}
    \country{Germany}
}

\renewcommand{\shortauthors}{Reeves and Simperl}

\begin{abstract}
Wikipedia is among the largest examples of collective intelligence on the Web with over 61 million articles covering over 320 languages. Although edited and maintained by an active workforce of human volunteers, Wikipedia is highly reliant on automated bots to fill gaps in its human workforce. As well as administrative and governance tasks, these bots also play a role in generating content, although to date such agents represent the smallest proportion of bots. While there has been considerable analysis of bots and their activity in Wikipedia, such work captures only automated agents that have been actively deployed to Wikipedia and fails to capture the methods that have been proposed to generate Wikipedia content in the wider literature. In this paper, we conduct a systematic literature review to explore how researchers have operationalised and evaluated automated content-generation agents for Wikipedia. We identify the scope of these generation methods, the techniques and models used, the source content used for generation and the evaluation methodologies which support generation processes. We also explore implications of our findings to CSCW, User Generated Content and Wikipedia, as well as research directions for future development. To the best of our knowledge, we are among the first to review the potential contributions of this understudied form of AI support for the Wikipedia community beyond the implementation of bots. 
\end{abstract}



\received{October 2024}
\received[revised]{April 2025}
\received[accepted]{August 2025}

\maketitle

\section{Introduction}

Wikipedia is one of the largest online examples of collective intelligence with over 61 million articles \cite{lewoniewski2023most} from 321 active languages \cite{moas2023automatic} as of 2023 making it the ``largest encyclopaedic work" ever created through human effort \cite{arroyo2020science}. Founded on the principle that anyone can create and edit content regardless of their expertise \cite{yarovoy2020assessing}, Wikipedia articles are developed through a highly collaborative process where editors collectively create, edit, refine and update articles either through modifying the articles themselves or through coordinated discussions on article `talk' pages \cite{crowston2020effects}.

It is perhaps unsurprising, then, that Wikipedia has long relied on bots and other forms of automation to manage the large quantity of content being created, edited and deleted each day \cite{halfaker2020ores}. As many as 16.5\% of edits in the English edition of Wikipedia are performed by bots and these bots make up a significant proportion of the most active Wikipedia editors \cite{traris2021leveraging}. Bots perform a wide variety of roles within Wikipedia, but most commonly they are employed to fix pages by correcting mistakes, linking pages or performing administrative tasks that would otherwise be time-consuming for the community \cite{zheng2019roles}. This use of bots for automating tedious or labour-intensive tasks within Wikipedia has become widely accepted by the user-base \cite{clement2015interacting}. 

Nevertheless, even with the assistance of bots, editing Wikipedia remains a labour-intensive process. Experienced Wikipedia editors are often expected to spend a significant proportion of their time on administrative and quality assurance tasks, leaving them little time to devote to the creation of new content, while newer editors may not post new articles without approval from more experienced editors \cite{ardati2024designing}. This reliance on human effort naturally leads to oversight in key editing tasks with over 50\% of Wikipedia articles lacking an improvement in quality over their life-cycle \cite{das2022quality}. The disparity between the availability of volunteer effort and the time required to curate and manage contributions is so extreme that the community has responded by increasing barriers to contribution and limiting the creation of new articles and content \cite{halfaker2020ores}.

To help fill this gap, Wikipedia has turned to content \textit{generation} bots which take responsibility for creating articles and content \cite{zheng2019roles}. Bot-generated content has a long history with the very first Wikipedia bot being a content generation bot intended to develop novel articles from US census results \cite{livingstone2016population}. Even today, some language versions of Wikipedia remain highly reliant on automatically generated content. Over 50\% of articles in the Serbo-Croatian and Vietnamese editions of Wikipedia are automatically generated, while a single user has automatically generated over a million articles (of disputed quality) in Egyptian Arabic Wikipedia \cite{alshahrani2023depth+}.

Yet in spite of this, these so-called ``generator'' bots remain relatively rare compared to other types of bot and are comparatively responsible for significantly fewer edits \cite{zheng2019roles}. Even under ORES, Wikipedia's initiative to develop key machine learning models, most pipelines focused on quality assurance with a single pipeline focused on drafting new articles typically through predicting article topics \cite{halfaker2020ores}\footnote{See also the pipeline repository on GitHub: \url{https://github.com/wikimedia/drafttopic}}. As Large Language Models and other AI systems improve, it is possible that in future, editors may be more readily able to make use of such systems to generate Wikipedia content. Nevertheless, it is not yet possible to do so effectively and responsibly -- on the contrary, Wikipedia's current policy on LLM usage strongly discourages writing article content\footnote{\url{https://en.wikipedia.org/wiki/Wikipedia:Large_language_models}}.

In this paper, we explore the state-of-the-art in research implementing automated generation models for Wikipedia. Through a structured review of research literature, we explore how researchers have operationalised AI models to support content creation tasks. We explore the scope of content generated through such models, the models used, input data for content generation and the methods used to evaluate such content. We further explore implications of this research for CSCW and User Generated Content (UGC) more broadly, as well as for Wikipedia specifically. We conclude with research directions for future work. By analysing how researchers design and implement content automation specifically within the context of Wikipedia, we offer insight into how AI might reshape notions of authorship in UGC communities, as well as the role of human users in such communities. 

\section{Background and Related Work}

\subsection{Automation in User-Generated Content Communities}

Previous CSCW work has explored and characterised the nature and role that bots play in peer-production, knowledge sharing and social online communities. \citet{wessel2018power} explored the role that bots play in Open Source Software projects, identifying 12 tasks performed by bots, largely with the role of supporting and notifying users rather than generating content. In a follow-up study, \citet{wessel2021don} interviewed users about their interactions with bots and found that users often characterised bots as distracting rather than the supportive force they were intended to be. \citet{berkel2024collaborating} analysed the use of bots and automation in the online crowdsourced mapping platform OpenStreetMap which they contrast with Wikipedia. Their findings suggested that automation is less disputed in Wikipedia and that OpenStreetMap users had a desire to democratise automative tools to make them more participatory and open. This prior body of work demonstrates the unintended consequences that automated tools can have and the range of responses that users may have to them. We build upon this work by exploring how model developers have accounted for -- and evaluated their approaches based on -- user feedback.

Another strand of research has explored the impact bots have on engagement and users within social networks and user generated content environments. Analyses of bot activity in social networks such as X/Twitter have identified a large proportion of content-generating bots which expose users to harmful or negative content \cite{mendoza2020bots, stella2018bots}. Nevertheless, earlier work has demonstrated that such bots can also promote engagement and content generation from human users, particularly in civic engagement and political contexts \cite{woolley2020social}. Conversely, \citet{safadi2024effect} analysed the impact of bots in the social media platform Reddit and found that even simple bots lowered user attachment to individual subreddit\footnote{Individual thematic Reddit communities which users can subscribe to and engage with. Examples are numerous but include r/Music or r/AskReddit} communities and reduced direct interaction between individual users. \citet{wei2024understanding} explored the impact of AI-generated content in the artistic social media platform Pixiv, finding that the introduction of AI-generated content led to a reduction in new user registrations and an increased homogeneity in content themes. \citet{seering2018moderator} analysed bot activity on the Twitch streaming platform and identified that bots can take on meaningful social roles in conversations.

Wikipedia is not a social network in the conventional sense and our analysis does not focus on the social elements of Wikipedia (such as Talk pages). Nevertheless, questions of how automation impact human engagement and of how content is curated and bias prevented are highly relevant to Wikipedia. We further address these questions by exploring the scope of automation and content generation as operationalised within the wider literature and the resulting implications for human agency and control of Wikipedia content.

\subsection{Automation in Wikipedia}

A significant focus within the Wikipedia research literature on automation has been understanding and identifying the activities performed by bots. \citet{zheng2019roles} analysed 1,601 bots within Wikipedia and characterised them based on their activities including name space used, edit frequency and software. The resulting typology identified 9 different types of bot with the most relevant to our work being the generator bot responsible for generating pages. This work focused only on bots which had been directly deployed to Wikipedia (and which were labelled as such). We build upon this by exploring how automated content processes have been proposed and implemented within the literature to capture the less visible and understudied issues of how automation \textit{could} be implemented in Wikipedia, how it is viewed by researchers and implications for the Wikipedia community as a whole.

While not directly analysing Wikipedia, \citet{hall2018bot} performed an analysis of edits within Wikidata\footnote{Wikimedia's knowledge-graph based database. For more details see section \ref{sec:Wikidata}} with a particular focus on identifying edits recorded as coming from human users but which were likely bot generated. To classify bots, they developed a framework of possible indicators including edit frequency and variation in edit comments. More relevant to our methods, \citet{ren2023did} performed a review of articles on Wikipedia editor behaviour, including an analysis of types of bot and their impact on editors, although without a specific focus on content generation bots. Both studies focused largely on characterising bots and their editing patterns, while we expand this to include automation processes beyond self-contained bots. We also analyse the content produced and consider the upstream data required to produce new content. In doing so, we aim to further characterise and distinguish the nature and role that automated agents could play in Wikipedia.

An additional focus within the literature has been understanding how editors and users perceive and interact with bots. \citet{clement2015interacting} explored talk-page interactions between users and bots and found that while users were generally accepting of bot activity, reactions were more extreme for `policing' bots designed to ensure users comply with Wikipedia's norms than they were for bots which simply carry out repetitive or laborious tasks. \citet{smith2020keeping} interviewed Wikipedia editors about the development of machine learning systems as part of the ORES initiative and found that while users were willing to accept machine learning models, they placed significant importance on the retention of overall control by human editors, as well as transparency of machine learning outputs. \citet{geiger2018goodbots} analysed bot-bot conflicts and interactions within Wikipedia, finding that in contrast to earlier characterisations of such interactions as negative, they in fact represent important collaborative workflows for content generation and improvement. While our work does not directly explore human-bot interaction, by considering the scope of automated content generation and the role of humans (particularly editors) in output evaluation, our review offers further perspectives on how indirect human-agent interaction may be operationalised by researchers and developers.

There have also been efforts to directly engage the Wikipedia community in developing and evaluating machine learning models. ORES\footnote{Objective Revision Evaluation System, since deprecated and replaced by LiftWing. See \url{https://www.mediawiki.org/wiki/ORES} for further details.} allowed users to create training datasets, commission models and evaluate pages using machine learning techniques \cite{halfaker2020ores}. More recently, \citet{kuo2024wikibench} presented WikiBench, a tool to empower the Wikipedia community in curating datasets for the development of AI-evaluation tools. By using a systematic literature review methodology, we aim to capture and further explore automated and semi-automated tools and technologies for the generation of Wikipedia content beyond the largely bot-based analyses that have been conducted previously. Based on the forms of engagement shown in this prior work, we review both the nature of model inputs and human evaluation of outputs to understand the role of the Wikipedia community and other human actors in model development and (where relevant) deployment.

\section{Wikipedia Content}

Wikipedia is divided into different categories of page known as \textit{namespaces}. Each namespace represents a different purpose or focus of a page and there is a corresponding talk page for each namespace\footnote{A summary of current name spaces can be seen at \url{https://en.wikipedia.org/wiki/Wikipedia:Namespace}} \cite{balestra2016motivational, mohamed2014identifying}. For example, namespace 0 details the main article content most commonly associated with Wikipedia, while namespace 2 represents editors' own user pages\textcolor{blue}{\footnote{\url{https://en.wikipedia.org/wiki/Wikipedia:User_pages}}}. Although we take a broad definition of Wikipedia content, we note that papers within our sample typically focus on the main workspace (number 0) with some relevance to the template (number 8) and category (14) namespaces.

\begin{figure}
    \centering
    \includegraphics[width=0.5\linewidth]{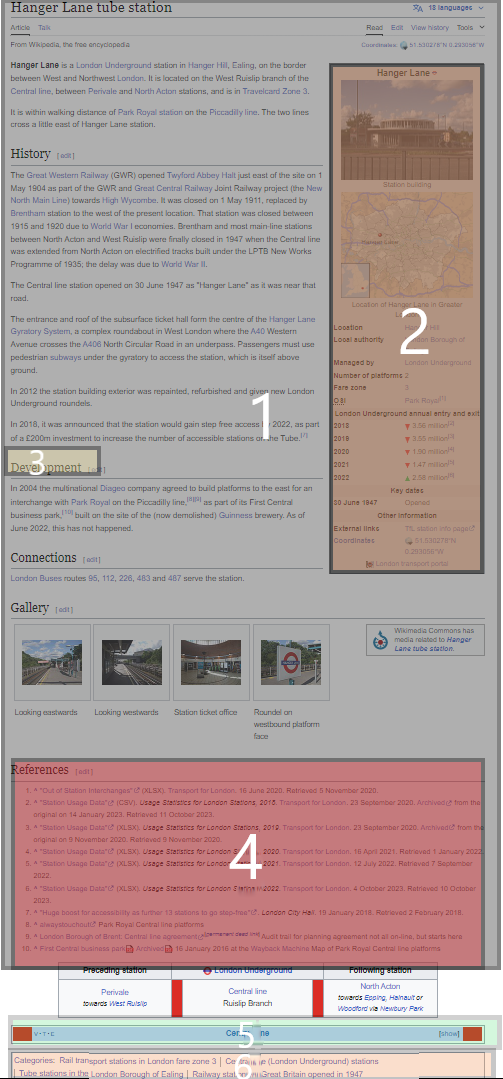}
    \caption{Outline of a typical Wikipedia article including 1 = full article, 2 = infobox, 3 = section headings, 4 = reference list, 5 = navbox, 6 = categories.}
    \label{fig:article-outline}
    \Description{Screenshot of the Wikipedia article ``Hanger Lane Tube Station'' with numeric labels to indicate full article (entirety of screenshot), infobox (upper right corner of screenshot), section headings (above image paragraphs), reference list (at end of main article text), navbox (penultimate structure in image) and categories (final structure in image).}
\end{figure}

Figure \ref{fig:article-outline} shows a typical layout for a Wikipedia article within the main namespace. In keeping with the papers within our analysis, we use the term \textit{article} to refer to the page in its entirety, with each text portion of the main article being referred to as a \textit{section} and each section being labelled with a \textit{section heading}. Many Wikipedia pages have an \textit{infobox} (labelled 2) which provides a summary of key concepts and statistics related to the subject of the article. For example, our example article is ``Hanger Lane Tube Station'' and the infobox features an image, location of the station, opening dates and passenger numbers among other details. Pages also feature a list of external references cited within the article and may feature a navbox\footnote{\Wikipediab}. This box is a table providing a structured collection of links to related articles in similar topical categories \cite{xie2017automatic}. Finally, pages are typically assigned to many different categories such as ``Central line (London Underground) stations'' and ``Railway stations in Great Britain opened in 1947" in our example. 

Although Wikipedia editors have taken responsibility for a number of initiatives to aid in structuring and scaffolding articles such as the creation of templates or how-to pages, volunteers retain a degree of freedom in writing pages which can be daunting for newcomers and inexperienced users \cite{piccardi2018structuring}. Where available, templates are also often rigid and language- and even topic-specific \cite{piccardi2018structuring}.

When defining ``content" we emphasise that the process of creating content extends beyond simply article text and may include creating article sections and headings or adding, populating and extending structural elements such as navboxes and categories. 

\section{Methodology}

The task of identifying automated agents within Wikipedia is not as simple as it may initially seem. While machine learning models have had some success in distinguishing fully automated agents, there remain outstanding questions around how to identify and distinguish human contributions augmented by semi-automated tools \cite{hall2018bot}. Given these difficulties, rather than navigating this distinction, we instead draw on a systematic literature review to identify the nature and applications of Wikipedia content generation models. 

\subsection{Literature Search}

We searched for articles with the word ``Wikipedia'' with a wildcard match ``generat*". In each of the databases we analysed, the * wildcard serves as a match for any number of characters. As a result, generat* would match generate, generated, generates, generation, generating and other variations of the term. Additionally, we include search terms for automat* with creat* or writ* to capture automatic creation and writing of article content.

Due to the relatively broad nature of our search terms, we chose to search only for papers where the keywords appeared in the abstract. Searches on the title field returned very few results while full text searches returned an unworkable number of articles which were likely to be largely irrelevant. Within Scopus\footnote{\url{https://www.scopus.com/}}, this was not possible and we used the closest equivalent (Title-Abstract-Key). Additionally, in the case of Scopus, we limited our search to only include journal articles (`ar') and conference papers (`cp') to avoid gathering irrelevant documents such as patents. We noted a level of inconsistency in some of the other databases around publication types (for example, `conference paper' would sometimes return workshop papers and sometimes not) and we therefore chose not to filter publication types beyond Scopus, instead considering publication type when evaluating studies for inclusion. The search returned a total of 4,909 articles across the databases (see Table \ref{tab:databases} below). After completing this process, we proceeded to the evaluation and selection stage.

\begin{table}
    \centering
    \begin{tabular}{|p{1.25in}|p{1.5in}|p{1in}|p{0.5in}|} \hline
        \textbf{Database} & \textbf{Search Terms} & \textbf{Search Fields} & \textbf{Results}\\ \hline
        ACM Digital Library & Abstract:("Wikipedia") AND Abstract:(Generat* OR (Automat* AND (Creat* OR Writ*))) & Abstract  & 464\\ \hline
        IEEE Explore & ("Abstract":"Wikipedia") AND (("Abstract":Generat*) OR (("Abstract":Automat*) AND ("Abstract":Creat*) OR ("Abstract":Writ*))) & Abstract & 322\\ \hline
        Sage & "Wikipedia" AND (Generat* OR (Automat* AND (Creat* OR Writ*))) & Abstract & 17 \\ \hline
        Scopus & ( TITLE-ABS-KEY ( "Wikipedia" ) AND TITLE-ABS-KEY ( generat* ) OR ( TITLE-ABS-KEY ( automat* AND ( creat* OR writ* ) ) ) ) AND ( LIMIT-TO ( DOCTYPE , "cp" ) OR LIMIT-TO ( DOCTYPE , "ar" ) )  & Abstract, Title, Keywords  & 2,706 \\ \hline
        Web of Science & AB=("Wikipedia" AND (Generat* OR (Automat* AND (Creat* OR Writ*)))) & Abstract  & 1,400 \\ \hline
    \end{tabular}
    \caption{Search terms, filter fields and number of articles for each queried database.}
    \label{tab:databases}
\end{table}

\subsection{Selection Process and Evaluation Criteria}

After conducting the search, we downloaded the full list of references from each database as an EndNote library. We then combined the lists and exported them as a tab-separated-values file. We manually evaluated the list of publications to identify duplicates and flagged 2,876 articles as duplicates.

Additionally, we made the decision to remove articles published before 2014 from the sample. While Wikipedia was first launched in 2001, it evolved rapidly in the years following its creation \cite{almeida2007evolution}. We noted that many of the earlier papers within our sample were potentially outdated either in terms of the methods and tools they used or served only to propose methods rather than implementing and evaluating these methods. Additionally, while comparing papers across time periods was somewhat difficult, we observed that at least some earlier studies appeared to be superseded or contradicted by more recent work. We chose 2014 as a cut-off point to cover only the most recent 10 years of publications (at the time of analysis).

When selecting papers for the sample, we evaluated papers according to the following criteria such that studies must:

\begin{itemize}
    \item \textbf{Be peer-reviewed} -- studies needed to be published in a peer-reviewed conference or journal. Workshop papers were included for peer-reviewed workshops only. Book chapters were removed due to potential variations in the review process and the lack of documentation of review processes.
    \item \textbf{Focus on automatic or semi-automatic generation of content} -- studies needed to generate content with minimum input from a user. Drafts and templates were considered to be in scope assuming that the template or draft would be produced without the user. Conversely, tools which would fill in a template based on answers from a user were considered out of scope.
    \item \textbf{Clearly focus on generating content for Wikipedia} -- studies which focused on more generic production of text or using Wikipedia content to generate content without relevance to Wikipedia were excluded.
    \item \textbf{Describe, implement and evaluate a content generation approach} -- theoretical publications or publications which proposed but did not implement such approaches were deemed to be out of scope.
\end{itemize}

Due to the reasonably large size of our sample, which included over 2,000 papers even after the removal of duplicates, we used an iterative approach to selecting papers. Firstly, we evaluated the papers by metadata only, removing papers from the sample based on year of publication or where the title was clearly unrelated to our analysis. We performed this analysis very cautiously and only removed papers where we could be entirely confident that the paper was not relevant. After this, we then analysed the abstracts for each of the remaining papers. At this stage, we evaluated papers against each of our evaluation criteria once again, removing only those papers which were clearly irrelevant. Finally, we then analysed the full text of each of the remaining papers. After evaluation, a total of 51 papers were selected. Table \ref{tab:exclusion} below shows the different exclusion criteria applied to the papers.

\begin{table}
    \centering
    \begin{tabular}{|c|c|}
        \hline
        \textbf{Category} & \textbf{Count} \\ \hline
        Duplicates & 2,876\\ \hline
        Incompatible with criteria & 1,593\\ \hline
        Non-English & 6\\ \hline
        Not Found & 3\\ \hline
        Not Peer-Reviewed & 19\\ \hline
        Prior to 2014 & 361\\ \hline
        Selected & 51\\ \hline
        \textbf{Total} & 4,909\\ \hline
    \end{tabular}
    \caption{Number of papers excluded or included based on exclusion criteria}
    \label{tab:exclusion}
\end{table}

\subsection{Positionality Statement}

Our aim is first and foremost to understand how researchers have operationalised content generation within Wikipedia. We draw a distinction between this and how bots are currently used. While generator bots exist in Wikipedia, we note that the bot policy\footnote{\url{https://en.wikipedia.org/wiki/Wikipedia:Bot_policy}} as currently defined poses particular challenges for large-scale generation of Wikipedia content due to the need to ensure approval for the generation of content. We also believe there are likely to be tensions and distinctions between the way in which Wikipedia's users view automated-generation and the way in which it is viewed (or proposed) by researchers.

We are ourselves users and at times editors of Wikipedia. Nevertheless, we are also artificial intelligence and CSCW researchers and we recognise that we make certain assumptions about the role that automated generation can -- or should -- play in communities like Wikipedia. For the purposes of this review, we have aimed not to make a judgement on the desirability of automated content generation for Wikipedia. Our goal with this review is not to explore the ethical and social impacts of such tools, particularly as the sampled literature does not discuss such matters. Instead, we focus on how such content generation has been proposed and framed by the research community and implications for Wikipedia, CSCW and User Generated Content more widely. 

We highlight, however, that there are wider questions around whether automated content generation at scale is a desirable goal in a community-maintained User Generation Community such as Wikipedia. While the source literature largely did not allow for exploration of this issue, we nonetheless recognise its importance and we discuss implications of our findings for the role of humans in AI-content generation within section \ref{sec:human-agency}. Our aim by highlighting these questions is to facilitate wider research and discussion by the CSCW and Wikipedia communities.

\subsection{Research Questions}

Our overall research question is \textit{How have researchers operationalised automated generation of Wikipedia content?} To address this, we identified a series of sub-questions:

\begin{enumerate}
    \item What is the scope of automatically-generated content models implemented within current literature?
    \item What methods and techniques have been used to generate content?
    \item What sources are used for content generation?
    \item How have these methods been evaluated?
\end{enumerate}

\section{RQ1 -- Scope of Automatically-Generated Content}

We begin our analysis by exploring how researchers conceptualise and operationalise automated content-generation methods within Wikipedia through a focus on the scope of the content which these methods generate. We identify each of the types of the content generated within the sampled studies, which we group according to the structure, format and role of such content. We conclude with Table \ref{tab:targ}, which summarises the type of content generated within each study.

\subsection{Article Sections} The largest proportion of studies within our sample focused on the task of generating one or more sections (with accompanying section headings) to either expand existing articles or create new ones. We distinguish this task from the task of generating article \textit{summaries} which did not generate novel content for a particular article and for which section headings were typically lacking.

The length and number of sections varied within the sampled studies. We also note a distinction in whether content was intended to stand alone as a `full article' or to expand existing pages. In such cases, the main distinction was largely the topical consistency and homogeneity of the generated content. For example, \citet{ambavi2019biogen} and \citet{duan2023wikipedia} both intended to create biographies for individuals with existing Wikipedia pages and used approaches intended to generate specific biographical sections and text. Conversely, in examples intended to generate full standalone articles, methods generated a more diverse selection of sections based on diverse topics. 

Despite the focus of some studies on generating `full' articles, however, we find little evidence of the generation of content beyond article sections and text (with the exception of embedding links to related articles). Although full Wikipedia articles may often feature infoboxes, navboxes or categories, none of the example content from studies within this grouping included these content types. For example, \citet{agarwal2023automatically} created Hindi articles for scientists with a minimum length of 500 words. The example generated article includes an image and a reference, but only one section and no infobox, navbox or categories. Conversely, \citet{zhu2020event} proposed an automated method for generation of articles, creating long-form articles comprising of multiple sections divided by topics, but again seemingly without infoboxes, navboxes or categories.

\subsection{Descriptions, Summaries and Abstracts}

We also observed studies aiming to generate abstracts or short descriptive summaries of articles. As \citet{sakota2023descartes} describe, although Wikipedia's guidelines advise that all articles should have a short description as a lead section\footnote{An introduction to an article covering its most important content. See \url{https://en.wikipedia.org/wiki/Wikipedia:Manual_of_Style/Lead_section}}, a significant number of articles have no such summary. While in practice, this process similarly entails producing article text content, we distinguish these studies from others due to their not requiring external references to produce content and the lack of novelty in the resulting text. This task can broadly be seen as a text summarisation task \cite{ta2023wikides}, but the structure and reliance on named entity links to related articles makes the process of producing article summaries in Wikipedia somewhat unique when compared to other contexts. As with other examples, the length and nature of these descriptions varied somewhat. For example, \citet{chisholm2017learning} produced one sentence summaries focused solely on biographical content, while in contrast, \citet{ta2023wikides} generated descriptions of varying length based on individual paragraphs. While these descriptions were typically intended to serve as an introduction to a larger article, \citet{kaffee2022using} proposed a method to generate summaries for article placeholders which feature facts and statistics from Wikidata\footnote{A collaborative knowledge-graph which provides machine readable data for Wikimedia projects \cite{piscopo2018models}}, but do not have long-form text from which the descriptions themselves can be produced.

\subsection{Sections Headings} \label{sec:headings} Wikipedia articles are structured into sections, which typically have a heading. While three of the papers in our sample focused on the task of generating section headings, the process, input and goals for generation varied significantly. \citet{field2020generative} detailed a method for generating section headings through clustering existing Wikipedia content with the goal of assigning section headings to improve existing Wikipedia content. \citet{piccardi2018structuring} propose a similar method to generate headings for articles, but in contrast with the method proposed by Field et al., this alternative method also generates section headings that can be used to bootstrap and create new Wikipedia articles. \citet{difallah2022crosslingual} aimed to identify and align section headings between languages on Wikipedia to allow easier creation of and/or translation into new languages for existing Wikipedia articles. 

\subsection{Images and Captions} One paper within the sampled literature describes the development of contextual images which can accompany Wikipedia articles. \citet{mitra2024generating} use a combination of LLMs and Text-to-Image models (TIMs) to generate images to accompany and provide additional context to Wikipedia articles. This is a challenging task as such articles typically follow a broader narrative covering a range of topics and details without generally focusing on visual descriptions or other contextual details (e.g., "style, background elements or mood") that models might usually rely on to generate imagery. While this was somewhat unique within our sample due to the focus on non-textual content, the study also described the process of producing descriptions for the images. We identified one further study in our sample with a focus on matching images to captions with the intention of generating image captions although we note that this could also be valuable for image generation models \cite{messina2024cascaded}. 

\subsection{Links} Within the sampled literature, we identified 5 papers which propose adding or expanding \textit{wikilinks} -- links to Wikipedia content. It should be noted, however, that links within Wikipedia content can appear in a variety of locations and are embedded within other forms of content. These links may appear in four key contexts within a given article including the summary infobox descriptions, named entities within the article text, the `see also' section at end of the article and the categorisation box at the bottom of the article content (which may also be duplicated on other articles linked within that category). The task of generating links is therefore an integral part of generating any other form of Wikipedia content, even if not explicitly considered within those models.

\subsection{Infoboxes} Some Wikipedia articles feature a summary `infobox' at the top of the page which summarises key facts and details about the article subject. Pages covering similar related topics such as animal species will typically follow a similar template across pages and languages. \citet{bhuiyan2015unsupervised} identified over 1800 infobox styles, noting that this caused difficulties for new and inexperienced editors who may not be able to recognise which the format, structure or content that best suits a given article. They proposed a method to automatically generate or otherwise identify templates from article content which was able to identify the appropriate template with an accuracy of between 60.97-78.58\% in experiments. A more fully automated method is proposed by \citet{saez2018automatically} who generate full infoboxes using content from Wikidata. While Wikipedia has pre-defined templates for creating infoboxes from Wikidata, the authors approach differs by eschewing these templates in favour of generating infoboxes from attribute value pairs. This method was trialled with 15 entities by generating 5 infoboxes per entity with a total of 25 attribute-value pairs per infobox. Finally, \citet{ambavi2019biogen} develop infoboxes for biographical articles. A common issue noted across these studies is the difficulty in identifying the optimum length and detail for an infobox.  

\subsection{Navboxes}

Navboxes are a table of links to related Wikipedia pages which typically appear at the end of an article \cite{xie2017automatic}. While the content of a navbox is functionally no different from a link, the structure of the navbox and the subjective nature of the link selection complicate the process of navbox generation. To generate a navbox, one must first identify the most relevant associated articles, but also filter those articles to avoid unnecessarily overloading the box with too many links or links of low relevance \cite{xie2017automatic}. This stands in contrast to the task of Wikilinking where the article text can serve as an input for the linked entities. Only one study within the sample (\cite{xie2017automatic}) explicitly explored the task of producing navboxes, but we note similarities to the task of creating links and the importance of such navboxes to full Wikipedia articles.

\subsection{References}

The process of adding references to Wikipedia articles is somewhat unique in that simply adding references to an article is not necessarily desirable in and of itself. References in Wikipedia articles are intended to support and provide evidence for individual claims. As a result, merely \textit{adding} references would be insufficient to improve an article and instead, the references would either need to be introduced alongside novel text or situated within the context of unsupported sections of the existing article text. Articles in our sample with this intended function therefore focused on suggesting references for use in producing content rather than augmenting articles. For example, \citet{joorabchi2020wp2cochrane} produced a tool that recommends articles from the Cochrane Library to produce content which can be used to update and improve medical Wikipedia articles. Adding references would also be necessarily for producing article sections and full-length articles. 

\subsection{Categories} 

Categories within Wikipedia play a predominantly structural role within Wikipedia with similar articles divided into common sections to support navigation. Categories have their own namespace, but also appear at the bottom of articles and therefore form part of the main namespace content. Only one study explicitly detailed the task of generating categories (\cite{pasca2020interpreting}) despite the fact that such categories would be required for full length articles. 

\begin{table}
    \centering
    \begin{tabular}{|c|>{\centering\arraybackslash}p{0.5\linewidth}|}
        \hline
        \textbf{Target} & \textbf{Sources}\\ \hline
        Section Text & \cite{agarwal2023automatically,zhu2020event,fan2022generating,liu2018generating,ranta2023multilingual,pochampally2016semi,duan2023wikipedia,banerjee2016wikiwrite, lu2020encycatalogrec,banerjee2015filling,prabhumoye2021focused,iv2022fruit,pariyar2014inconsistency,kaffee2018mind,kaffee2022using,banerjee2015wikikreator,sharevski2020wikipediabot,chandraseta2019composing,ahmeti2017updating,pfundner2015utilizing,alegriawikipedia, field2020generative,difallah2022crosslingual,piccardi2018structuring,chen2021wikitablet,rong2023fufaction,chen2019enhancing,banerjee2015bwikikreator,zhu2021twag}\\ \hline
        Article Summary & \cite{sakota2023descartes,chisholm2017learning,kaffee2018learning,zhu2020event,ta2023wikides,gao2021biogen}\\ \hline
        Images & \cite{mitra2024generating}\\ \hline
        Image Captions & \cite{mitra2024generating,messina2024cascaded}\\ \hline
        Infobox & \cite{bhuiyan2015unsupervised,saez2018automatically,ambavi2019biogen,yang2019cross,zhang2014wiicluster}\\ \hline
        Links & \cite{ikikat2015automatic,yang2019cross,schwarzer2016evaluating,scharpf2021fast,pang2022multi,wang2016error}\\ \hline
        Labels & \cite{asthana2021automatically} \\ \hline
        Navbox & \cite{xie2017automatic}\\ \hline
        References & \cite{joorabchi2020wp2cochrane}\\ \hline
        Categories & \cite{pasca2020interpreting}\\ \hline
    \end{tabular}
    \caption{Target content generated by sampled literature}
    \label{tab:targ}
\end{table}

\section{RQ2 -- Content Generation Methods and Techniques}

In this section we explore RQ 2 - What methods and techniques have been used to generate content? We group models into one of five broad categories: template and rule-based approaches, topic model approaches, graph-based approaches, sequence-to-sequence models and transformer-based approaches. Although partially linked with the scope of the generated content, models and techniques were selected and employed independently of the content generated.

\subsection{Template and Rule-Based Approaches}

Template-based approaches describe the generation of content by extracting data and inserting it into pre-defined formats. In the context of Wikipedia, these formats may include template sentences \cite{agarwal2023automatically} or more structured content such as infobox templates \cite{ambavi2019biogen} or placeholder tables \cite{kaffee2022using}. Rule-based approaches, conversely, describe the generation of content through pre-set logic or rules which can be used to generate sentences at length. In the context of Wikipedia, this might involve the generation of content into paragraphs and logically sequential sections based on predefined expectations of how these sections might fit together \cite{agarwal2023automatically}. While these approaches differ, we group them based on similarities in their requirements and within their strengths and weaknesses. Both of these methods require that researchers pre-define how content should be generated.

\textbf{Strengths} Template- and rule-based approaches offer strong control over the structure, factuality, and stylistic consistency of generated text, making them particularly well-suited to high-stakes domains like Wikipedia article generation. In the approach used by \citet{agarwal2023automatically}, the use of pre-defined sentences and rules offers highly predictable and controlled generation of text. Rule-based systems have been shown to help minimise risks of generating text that is nonsensical either syntactically \cite{gatt2018survey} or semantically \cite{abdelghafour2024hallucination}.

\textbf{Weaknesses and Challenges} Despite their reliability, template-based systems require considerable labour on the part of developers and they do not scale well to contexts where significant diversity is required in produced text \cite{gatt2018survey}. Rule based approaches are brittle and perform poorly or fail to generate outputs when encountering contexts that do not fit their rules \cite{clarke2023rule}. We note that both approaches are likely to scale poorly across domains or particularly languages due to the need to develop novel rules and templates. For example, while \citet{agarwal2023automatically} were able to produce templates for Hindi, these would need to be completely rewritten to suit other languages. 

\subsection{Topic Model Approaches}
In contrast to the more rigid rule/template-based approaches, topic model approaches allow for text to be generated in a template-based or structured format without the need for first defining rules, by instead a topical model and hierarchy from other, similar articles. For example, to produce a full article, \citet{zhu2020event} first induced a template of subtopics (a `topic tree') from existing articles and then generated text for each subtopic. Similarly, \citet{zhu2021twag} used predicted topics to guide abstract generation. Topic modelling was also used to scaffold or classify text without explicitly generating it -- for example, \citet{piccardi2018structuring} used topic modelling to identify the topics of Wikipedia sections, which could then be paired with a recommender algorithm to produce appropriate sections to help editors expand lower quality articles. 

\textbf{Strengths} Inferring topic models from articles enables the production of templates without the need to dictate rules and templates, lowering the labour required for developers \cite{gatt2018survey}. Topic modelling approaches are easier to train than more complex deep learning or neural network models and do not require such large training sets \cite{potapenko2017interpretable}. This makes these approaches more suited to more moderately sized corpora such as topic-specific articles and sections \cite{zhu2020event}.

\textbf{Weaknesses and Challenges} Standard topic modelling approaches struggle with larger vocabulary sizes, often necessitating that the most and least frequent terms in a set of documents are removed from the model \cite{dieng2020topic}. Moreover, while strong for classification and clustering, topic modelling in itself is less effective than more abstractive methods in generating coherent or creative text \citep{yu2022survey}. Topic modelling is difficult to apply to culture-specific content where the diversity of topics and languages can be challenging for existing models, particularly in some language editions such as Serbian where the diversity of topics is very high \cite{markoski2021cultural}.

\subsection{Graph-Based Approaches}

Graph-based approaches leverage the relational structure of connections between Wikipedia entities (e.g., articles and topics) to construct graph representations, which are then evaluated using inference, ranking and clustering approaches to generate content based on graph topology. For example, \citet{wang2016error} used \textit{LinkRank}, a graph-based ranking algorithm which detects and corrects erroneous hyperlinks by weighting links between entities based on fit. Alternatively, \citet{lu2020encycatalogrec} recommended categories (and by extension, section content) by constructing a graph-based workflow, predicting the weight associated with given nodes (articles and associated categories) to recommend the most appropriate categories associated with each article. \citet{yang2019cross} identified cross-lingual consistency of Wikipedia content by embedding statements as knowledge-graph triples and used random walk-sampling \footnote{A process for identifying randomly selected paths through an environment such as a graph. See \cite{xia2019random}} to model local and global structures within the graph to identify entity alignment.

\textbf{Strengths} Graph structures represent and encode relational assumptions allowing for explicit identification of data provenance and to ultimately improve the trustworthiness and validity of collaborative knowledge-sharing projects \cite{piscopo2017provenance}. Graph methods are also able to capture both direct and indirect relationships between nodes to capture similarities between articles that may not be directly linked. For example, \citet{yang2019cross} aligned entities by considering both local (i.e., more directly linked) and global (i.e., less directly linked) relations. 

\textbf{Weaknesses and Challenges} Sparsity in the structure of a graph or noise artifacts such as incorrectly defined links within Wikipedia can distort rankings and therefore model performance \cite{wang2021graph}. Articles with few links or with limited metadata are unlikely to fit within the graph without additional contextual details \cite{lu2020encycatalogrec}. Knowledge-graph representations of data also scale poorly and become computationally expensive as graph sizes increase \cite{ji2021survey}.

\subsection{Sequence-to-Sequence Models}

Sequence-to-Sequence models (StS) function by first encoding an input, then performing a transformation to decode it into a new text sequence. StS models were commonly used within our sampled studies to generate Wikipedia text from sources, particularly where text was lengthy or where the semantic or sequential relations present in the source text needed to be maintained in the generated content. For example, \citet{chisholm2017learning} used a sequence-to-sequence approach to generate short one-sentence biographies noting that contextual information from source data could often be lost due to a failure to encode relationships between terms. A similar approach was used by \citet{kaffee2018mind} who similarly aimed to ensure that summarised content accurately represented source data.

\textbf{Strengths} StS models learn to produce fluent sequential text that is able to integrate external information while producing highly novel text \cite{shi2021neural}. Generated content is not directly reliant on the exact terms used in the source document (i.e., it is abstractive rather than extractive). This approach is therefore particularly suited to generating Wikipedia content where copyright material may be involved as such material is likely to be removed if directly quoted \cite{banerjee2016wikiwrite}. 

\textbf{Weakness and Challenges} Neural-network based StS methods require extremely large datasets and can perform poorly at recognising less common patterns in source text \cite{lake2018generalization}. Additionally, performance is much lower for very long sequences of data \cite{lake2018generalization, zhao2020rnn} which may make them a poor-fit for article-level Wikipedia generation tasks. Neural-network based StS methods have also increasingly been shown to be at risk of \textit{hallucination} \cite{zhou2021detecting}, a phenomenon whereby the model generates text that is either incorrect or entirely unsupported by the training data and inputs \cite{ji2023survey}. 

\subsection{Transformer-Based Approaches}

Transformers are similarly a type of sequence-to-sequence model consisting of an encoder and a decoder. However, in contrast to other such models, transformers include \textit{cross-attention} mechanisms which allow decoder outputs to be passed back to the encoder, while encoder outputs can be fed back into decoder outputs \cite{lin2022survey}. \citet{sakota2023descartes} used \textit{mBART} (a multilingual transformer) for abstractive summarisation of descriptions from article content. Similarly, \citet{ta2023wikides} used a transformer-based approach to generate descriptions which was combined with a ranking algorithm to better select the most appropriate content for descriptions. \citet{zhu2020event} proposed a transformer-based approach to summarise content from topically relevant Wikipedia articles to bootstrap new articles on emergent events.

\textbf{Strengths} Transformer-based approaches are able to process larger sequences simultaneously, reducing training time and allowing them to scale better as inputs grow \cite{vaswani2017attention}. Indeed, transformer-based models are able to use larger training sets than other sequence-to-sequence methods and show higher performance with sufficiently large inputs \cite{lin2022survey}. They are able to account for long-range dependencies between two tokens in a sequence (e.g., relationships between words that are far apart) \cite{wu2021representing}. These strengths make them particularly suitable for working with full-length and longer Wikipedia articles as well as larger contexts such as cross-topic or cross-language datasets \cite{liu2018generating}. 

\textbf{Weaknesses and Challenges} Transformers require significantly more memory and are more computationally intense, particularly for longer sequences of text and this is compounded by the requirement to fine-tune and pretrain each model to achieve maximum performance \cite{pfeiffer2020adapterhub}. Transformers are also at particular risk of hallucination through the generation of fluent but inaccurate text \cite{maynez2020faithfulness}, which is a concern in a context like Wikipedia where verifiability is key. Although performance is high, transformers are also inefficient, particularly as few contexts require full implementation of cross-attention mechanisms\cite{lin2022survey}. These factors pose challenges to working with very large amounts of data, which may pose challenges for Wikipedia. At the time of writing there are almost 7 million articles in English alone\footnote{\url{https://en.wikipedia.org/wiki/Wikipedia:Size_of_Wikipedia}}.

\section{RQ3 -- Sources for Content Generation}

Wikipedia requires that statements in articles are factually accurate and that they contain citations to trustworthy external resources\footnote{For more details, see \url{https://en.wikipedia.org/wiki/Wikipedia:Reliable_sources} and \url{https://en.wikipedia.org/wiki/Wikipedia:Verifiability}}. As a result, many of the sources within our sample relied on external datasets and sources to generate content. In this section, we explore the sources used to generate materials, as well as the benefits and limitations associated with each source and the implications of this usage for Wikipedia. This section addresses research question 3 -- what sources are used for content generation?

\subsection{Specific Article}

The simplest studies within our sample did not rely on external source materials beyond the content of the article to be modified. This is not to suggest that the models used did not require training with other types of data, but rather that once trained, the functional model required only a single article as an input. This group of studies primarily focus on structural content and features such as links. An example of this is the work of \citet{ta2023wikides} which generated summaries of Wikipedia articles. While the authors trained their model using multiple Wikipedia articles, the ultimate form of the model was able to generate summaries from an input of a single article. Similarly, \citet{bhuiyan2015unsupervised} used 4.72 million articles to train and refine their model, but after evaluation and refinement, the model was able to suggest infobox templates for an article based on the content of that article alone. The lack of reliance on external materials or other content beyond the training data allows such approaches to be applied to a wide range of Wikipedia articles, but such an approach is only suitable for a limited number of use cases.  Although theoretically some other models were able to generate content based on an article alone (e.g., \cite{duan2023wikipedia}) in practice, such approaches would require information from other articles and/or external sources to comply with Wikipedia's regulations.

\subsection{Existing Wikipedia Content}

By far the largest proportion of the sampled studies relied heavily on existing Wikipedia content to generate new content and/or otherwise expand Wikipedia. Most commonly, studies relying solely or predominantly on existing Wikipedia content did not aim to generate long-form content, but rather to provide structural content to augment and improve articles. This content would take the form of section headings and infoboxes or through connections to related articles through internal links and navboxes. 

While in some cases, studies aimed to produce novel textual content or article sections, a key weakness of this approach was the availability of external reference materials. Since Wikipedia content must be factually accurate and supported by external references, novel content produced through this method may not be suitable for Wikipedia. Solutions to this challenge within the literature typically relied on contributions from human participants. For example, \citet{chen2021wikitablet} produce article text and sections based on tables of data within Wikipedia with these tables being produced by editors. 

We did observe one study that proposed generating `novel' Wikipedia content from existing articles. In the approach proposed by \citet{zhu2020event}, although the overall article topic (and key factual information) is sourced from Web sources, individual sections and paragraphs are synthesised based on articles linked to relevant topics for that article. We note, however, that it is not clear how referencing would -- or should -- function in such a context and Wikipedia's policies make it clear that Wikipedia is not a reliable source for Wikipedia content\footnote{see \Wikipedia}. Another approach proposed by \citet{ranta2023multilingual} entailed using newly developed Wikipedia content to produce multilingual content. However, this appears to only have been a proposal at the time the article was published and it is not clear how this might function in practice.

\subsection{Wikidata}
\label{sec:Wikidata}

Wikidata is a collaborative knowledge graph developed through contributions from human editors and bots in a similar style to Wikipedia \cite{piscopo2018models}. Wikidata consists of triples where a subject entity (i.e., the topic of a page) is associated with objects (values or other subjects within Wikidata) through predicates (properties of the subject) \cite{turki2019wikidata}. For example, the Wikidata subject Earth\footnote{\url{https://www.wikidata.org/wiki/Q2}} has the predicate ``instance of'' with the object value ``terrestrial planet''\footnote{\url{https://www.wikidata.org/wiki/Q128207}}, itself a subject within Wikidata. Wikidata relies on external resources to provide evidence for its subject-object relations with approximately 10\% of Wikidata entities featuring such links \cite{haller2022analysis}.

Content from Wikidata was used to generate a variety of content within the sampled studies including full and entire articles \cite{agarwal2023automatically}, infoboxes \cite{saez2018automatically} and article summaries \cite{chisholm2017learning, kaffee2018learning}. Wikidata had several advantages compared to other sources including pre-existing evidence of statements and relations, the machine-readable nature of the content and the range of languages represented within Wikidata. Conversely, Wikidata shares some of the quality and coverage issues associated with Wikipedia including imbalances in multilingual content \cite{amaral2021assessing}, gender and other biases \cite{zhang2021quantifying} and a long tail of low-quality entities \cite{chen2023knowledge}.

\subsection{Web Content}

The final group of studies used content from the web to support the generation of Wikipedia content. The majority of studies within this group relied on web searches to gather additional data about a subject which could then be modified to suit the Wikipedia format. While other sources used specific resources, approaches to this were highly variable. Not all web sources are suitable for Wikipedia\footnote{see for example \url{https://en.wikipedia.org/wiki/Wikipedia:Using_sources}} and as a result, simply finding a source through a search does not guarantee that the source will be appropriate for use in Wikipedia. \citet{pochampally2016semi} approached this challenge through the use of a weighted dictionary of textual web sources which was selected according to known suitability for Wikipedia. Conversely, other studies relied on specific sources to the exclusion of all others with such sources generally including other crowdsourced and collective intelligence platforms such as IMDB \cite{ambavi2019biogen} or DBPedia \cite{prabhumoye2021focused}, as well as social networking systems Flickr and Twitter (now known as X) \cite{laere2014georeferencing}. 

This approach is arguably essential for some types of task. Approaches which aim to address gaps within content in Wikipedia or to capture phenomenon from outside of Wikipedia cannot rely on Wikipedia content to fill those gaps. Yet, we also note that these approaches are very dependent on the available web search results. Due to limitations in terms of scaling and run-time, the approaches at times rely on as few as 20 search results (e.g., \cite{zhu2020event}) and it is therefore feasible that these results may not be appropriate for Wikipedia. We also believe that there exist trade-offs between the needs of the methods used and the assurance of quality sources. 

\section{RQ4 -- Evaluation Methods}

In this section, we explore RQ 4 - How have these methods been evaluated? Within the sampled literature, we identified four broad categories of method evaluation: computational evaluation, human evaluation without editors, human evaluation with editors and deployment of content to Wikipedia. We outline each of these approaches with examples from the literature and explore the goals and challenges of these evaluation approaches.

\subsection{Computational Evaluation}

The largest group of 44 studies used computational and quantitative evaluation methods without human oversight. Metrics used within this group focused largely on the linguistic similarity between the generated and source content. For example, \citet{chisholm2017learning} used BLEU scores to identify how similar their generated biographical summaries are to the source inputs, while \citet{zhu2020event} and \citet{chen2021wikitablet} used ROUGE to evaluate the similarity in summary text with source articles or structures such as tables. More broadly, metrics focused on precision and recall as indicators of method performance and error-rates.

While these methods avoid the need to recruit and manage users, they fail to adequately capture the quality and suitable of text for Wikipedia. Despite comparing \textit{similarity} between sources, this similarity is largely semantic or term-based and does not serve to capture the stylistic similarity or adherence to Wikipedia's content policies\footnote{See \url{https://en.wikipedia.org/wiki/Wikipedia:List_of_policies}}.

\subsection{Human Evaluation (Non-Editor)}

To facilitate qualitative evaluation and evaluation beyond that provided with computational metrics, 10 studies engaged humans without Wikipedia experience (or who were not recruited based on experience) to review or refine Wikipedia content. This group typically consisted of crowdworkers or students. While this form of evaluation offered greater insight than simply computational evaluation, it was not intended to understand the suitability of content for Wikipedia and relied on subjective measures such as user preference expressed as a likert score (e.g., \cite{saez2018automatically}).

A second class of evaluation within this group was manual evaluation by experts who were not necessarily editors. This method was used by \citet{pochampally2016semi}, who monitored the number and extent (as a percentage) of the edits that participants needed to make before automated outputs would be suitable for Wikipedia. In contrast to the automated and non-expert evaluation methods described, this did touch on suitability of text for Wikipedia by editing text until it was suitable to be launched to Wikipedia. However, it is not clear how this was ascertained and the evaluation of suitability appears largely stylistic and linguistic rather than specifically being based on adherence to the Wikipedia content policies. A total of 3 studies used expert review of outputs while a further 3 studies did not specify the nature of their reviewers.

\subsection{Editor Evaluation}

Only 4 studies employed evaluation by experienced or active Wikipedia editors to evaluate whether content was likely to be suitable for Wikipedia. \citet{piccardi2018structuring} asked experienced editors to evaluate their article suggestions as automated methods failed to adequately account for differences in meaning and syntax. While the authors also used crowdworkers with no Wikipedia experience, the two groups differed significantly due to the editors accounting not only for the linguistic suitability of suggestions, but also based on their own views of what a suitable section would be. \citet{kaffee2018learning} recruited editors to evaluate model selections and also monitored whether editors re-used model-generated content in their own work. This approach was repeated in two follow-up studies (\cite{kaffee2018mind, kaffee2022using}).

In contrast to other forms of human evaluation, this method was most likely to indicate whether content was suitable for -- and could be launched to and survive within -- Wikipedia. Nevertheless, evaluation within this group was small scale with small groups of editors individually (and independently) determining the suitability of content. Since Wikipedia is collaboratively-edited, no one editor's opinion is necessarily worth more than any other's and this method therefore does not guarantee the suitability of content.

\subsection{Deployment to Wikipedia}

Wikipedia content is typically improved, edited and replaced over time through the collaboration and efforts of volunteers (and potentially bots). Human refinement of outputs, then, is an implicit part of Wikipedia. 7 studies within our sample chose to take advantage of this process by deploying generated content to Wikipedia and monitoring edits. \citet{banerjee2015filling}, for example, deployed a generated article to Wikipedia and used the edits and feedback from users to refine content generation processes for two follow up studies (\cite{banerjee2015wikikreator, banerjee2016wikiwrite}). \citet{kaffee2018learning} similarly used the deployment of content to evaluate generation methods by exploring the re-use of generated text by editors. 

While this method arguably provides the greatest insight into the suitability of content for Wikipedia, it is not without its risks. As \citet{banerjee2015wikikreator} highlight, content can be easily removed for being unsuitable, particularly where it contains extracted text from copyright materials. Even if content is not removed, there is no guarantee that deployed content will receive edits. We also note the lack of a formal framework within the sampled literature for the formal evaluation of edits made to an article.  We therefore note that this method is likely to be better suited to larger forms of content (e.g., full articles) for which there is more scope for editing by Wikipedia users.

\section{Discussion}
\label{sec:disc}

In this section, we begin by revisiting our research questions with a summary of our findings. We then discuss the implications of our findings for CSCW and for User Generated Content more broadly, before concluding with a discussion of implications of our findings for Wikipedia.

\subsection{Research Questions}

\subsubsection{RQ1 -- What is the scope of automatically-generated content models implemented within current literature?}

We found no evidence of any approach that generated all of the content possible within the main article namespace. Although we did identify studies which self-described as generating full articles such as \cite{agarwal2023automatically,pochampally2016semi,zhu2020event}, these generated articles consisted solely of article and section headings, section text, links and in the case of \cite{pochampally2016semi}, an image. This leaves substantial scope for additional forms of content such as infoboxes, navboxes, references or `see more' links. In all other cases articles focused exclusively or almost exclusively on one form of content to the exclusion of others. It is not possible to identify whether this was linked to the research itself (e.g., an interest only in text generation), a necessity from the selected techniques and methods or linked to perceptions of the needs of Wikipedia. Regardless, we note that it would be possible to generate the full scope of an article's content by combining methods within the sample. 

\subsubsection{RQ2 -- What methods and techniques have been used to generate content?}

We identified a range of methods and techniques for the generation of Wikipedia content. More recent methods have tended to use sequence-to-sequence approaches (particularly transformers) for generating content, regardless of the intended scope of the generated content. Nevertheless, we do note outliers (e.g., \citet{agarwal2023automatically}) where more recent research has used more traditional rule- and template-based generation methods. We note that sequence-to-sequence and transformer-based approaches are likely to be particularly well suited to Wikipedia due to their ability to generate novel and paraphrased text. As described by \citet{banerjee2016wikiwrite}, the failure of more extractive methods to process copyright material has led to content generated with such methods being removed. Nevertheless, we believe this is an area that is likely to see substantial change as novel technologies and the availability of computational resources influence the feasibility and efficiency of distinct content generation methods.

\subsubsection{RQ3 - What sources are used to generate for content generation?}

While a variety of sources were used to generate content within the sampled studies, the most common source was Wikipedia itself. This is not necessarily an issue for some forms of content: we note that summaries, for example, should derive from the article they aim to support while other types of content such as lists and categories must, by necessity, use Wikipedia as a source. Yet, we also identified studies which used Wikipedia as a source to generate ostensibly new content. This raises concerns about verifiability as Wikipedia content cannot and should not be used as reference material or evidence for Wikipedia articles\footnote{see \Wikipedia}. We also note a variety of content drawn from the wider Web, often without specific attempts to ensure the verifiability of suitability of this content. We discuss these issues further in section \ref{sec:verifiability}.

\subsubsection{RQ4 - How have these methods been evaluated?}

The majority of studies within our sample focused exclusively on computational metrics and benchmarks to evaluate the generated content in terms of fluency rather than in terms of suitability for Wikipedia. While this is an important consideration when writing Wikipedia articles\footnote{See for example \url{https://en.wikipedia.org/wiki/Wikipedia:Policies_and_guidelines} which discusses the importance of readability.}, such considerations are not the sole factor in determining the likelihood that Wikipedia contributions will be accepted. Indeed, Wikipedia editors regularly make changes to articles such as adding clarification, copy editing, refactoring or even removing articles to improve content which is not inherently inappropriate or poorly written \cite{yang2017identifying}. While we recognise that the use of such benchmarks is important, we nonetheless caution that evaluation through benchmarks alone is unlikely to capture the complex and sometimes subtle nuances of article evaluation as conducted by the Wikipedia community. 

Engagement with Wikipedia users was comparatively rare. Only 4 studies engaged editors directly as part of the experimental design and analysis process, while 7 studies deployed content to Wikipedia and monitored changes to content, but did not directly communicate or engage with editors. We believe this is an important area for future work, particularly due to the relatively unstructured evaluation used in cases where content was deployed to Wikipedia.

\subsection{Implications for CSCW}

\subsubsection{Participation and Equality}

In ostensibly open knowledge-gathering communities such as Wikipedia, regardless of the open nature of the community there is significant scope for epistemic injustice which removes -- or renders invisible -- submissions from some individuals or groups \cite{ajmani2024whose}. We observe a risk of participation inequalities between human and bot users, particularly across languages. For example, while the methods proposed by \citet{kaffee2022using} or \citet{agarwal2023automatically} promote the availability of information in diverse languages, these approaches also serve to replace effort by speakers of those languages. At their worst, if applied to minority languages with a group of small, largely monolingual editors, such approaches could restrict or remove the opportunity for editors to perform creative work or to allow editors from these languages to influence the topical and informational focuses of articles.

We are not the first to raise concerns about the impact that AI-driven technologies may have in this space. Such inequality has already been well documented in a range of algorithmically-mediated platforms within and beyond the User Generated Content domain \cite{lutz2019digital}. Nevertheless, we believe that these trade-offs between the nature of platforms such as Wikipedia as an information gathering and a peer-production platform are likely to raise tensions as the accessibility and role of automated content generation grows through the availability of tools such as LLMs.  

\subsubsection{Human Agency and Creativity}
\label{sec:human-agency}

We also believe this review raises important questions about what the goal of using AI in human-generated content should be. What should be the scale of automated content generation? Is it desirable -- or even necessary -- to restrict the scale of generated content to leave gaps or opportunities for human input? While the studies within our sample did not generate the full scope of article content, they also did not provide sufficient context to understand whether this was a conscious choice and if so, whether it was made on technical, ethical or Wikipedia-specific grounds. These tensions have already been observed and identified in Wikipedia (see for example \cite{geiger2018goodbots}), but we believe they are of equal -- if not greater -- importance to other User Generated Content contexts which may lack the community-driven oversight of Wikipedia and where there is therefore greater opportunity for automated agents to co-opt human roles. 

We also feel there are questions raised about how we should balance the role of automated agents in supporting and nurturing human creativity without overshadowing or replacing it. Beyond the automated generation of content, studies within our sample raised the prospect of using such tools to support learning and improvement among Wikipedia editors or to promote the consistency of Wikipedia articles \cite{piccardi2018structuring, zhu2020event}. Yet, other perspectives have placed value on the act of making errors \cite{liu2018power} and the importance of struggle and challenge for characterising \textit{human} creativity \cite{lockhart2024creativity}. There is a danger, then, that rather than \textit{helping} editors to learn to edit, they instead \textit{prevent} editors from doing so by removing opportunities for feedback or making them reliant on the use of such tools.

\subsubsection{Generalisability} Although our analysis focused on Wikipedia, we note that the models and techniques could be applicable to other types of User Generated Content. While we observed some types of content that were likely only suited to Wikipedia due to their format (e.g., navbox and infobox templates) or due to referencing Wikipedia specific concepts (e.g., categories), on the whole the general nature of the content and the broad focus on section text is likely to be applicable to a range of User Generated Content contexts. However, we note potential barriers to this generalisability in terms of the techniques and source content used. The reliance on rule- and template-based approaches as used by \citet{agarwal2023automatically} necessitates substantial effort to extend such tasks to other languages and potentially even topical contexts. The reliance of many studies on transformer-based methods and the associated requirements for large training datasets also makes them poorly suited to languages with less Wikipedia content such as Tigrinya, which has just 336 articles and is the 7th smallest Wikipedia edition at the time of writing\footnote{\url{https://en.wikipedia.org/wiki/List_of_Wikipedias}}. 

\subsection{Implications for Wikipedia}

\subsubsection{Attribution} Wikipedia's policies on AI-generated content currently only require that content is attributed to an AI if the process is fully automated -- i.e., that there is no human oversight in the process. Yet, our analysis suggests the possibility of fully (or almost fully) AI-generated content that would not be recorded as AI-produced. One example of this is the content generation process described by \citet{agarwal2023automatically} where the content recorded to Wikipedia was recorded from an anonymous user account, despite the limited human input into the resulting model output. We believe this is likely to worsen with the advent of generative AI, with human users able to use such tools to assist or even fully generate content for them. While Wikipedia's current policies strongly advise against full generation of content using LLMs, they do allow for editors to use LLMs to support article creation under specific circumstances\footnote{\url{en.wikipedia.org/wiki/Wikipedia:Large_language_models}}. We therefore believe that there is increasingly a need for new content generation rules to recognise the contributions of AI agents. This could be achieved through a message on article talk pages documenting the extent and nature of AI contributions. 

Moreover, we also note epistemic implications around what counts as novel knowledge and what is a verifiable source. Many of the sampled models relied predominantly on intrinsic knowledge drawn from Wikipedia or from other Wikimedia projects such as Wikidata. However, can or should Wikipedia articles be considered a verifiable source and how far removed may content be from the original verified source before it should no longer be considered verifiable? In examples such as \citet{zhu2020event}, as a model potentially generates articles related to a specific topic, so the likelihood increases that at least some of those articles will influence subsequently generated content. 

\subsubsection{Verifiability} \label{sec:verifiability}Any information contained within the main namespace of Wikipedia is required to be verifiable\footnote{\url{https://en.wikipedia.org/wiki/Wikipedia:Verifiability}}. Yet, we believe there is a risk of unsuitable or even completely unverifiable content within the sampled studies due to reliance on unevaluated Web content (e.g., \cite{chisholm2017learning}) or new content synthesised predominantly from other Wikipedia articles rather than external sources (e.g., \cite{zhu2020event}). There are, however, resources within the sample that allow for this to be addressed such as a focus on pre-defined dictionaries of suitable web sources (e.g., \cite{pochampally2016semi}) or resources such as WikiData which encode facts and information about entities as extracted from verifiable sources\footnote{\url{https://www.wikidata.org/wiki/Wikidata:Verifiability}}. We suggest future methods should utilise such resources and focus specifically on questions of verifiability as it is ultimately one of the factors which best distinguishes Wikipedia content generation from other Natural Language Generation tasks. 

\subsubsection{Researcher-Community Interaction} Currently, we note mismatches between the way in which automation of Wikipedia is operationalised by researchers and the way it appears to be perceived by editors. This is perhaps best exemplified by the example of \citet{zhu2020event} who describe the motivation behind their approach as: ``writing a Wikipedia document manually can be time-consuming and difficult, so automating the writing process is a valuable research topic." While this may be true, automating the writing process is not necessarily seen as a valuable use of automation by the Wikipedia community. Indeed, Wikipedia communities have often resisted attempts to fully automate content generation and have shown a preference for retaining control and oversight of the content generation process \cite{smith2020keeping}. As well as manifesting in the articulation of motivations, we also observe these tensions in the failure to engage the Wikipedia community in evaluation and in the general failure to integrate methods or their outputs into Wikipedia.

On the one hand, this is not fully surprising as our analysis focused explicitly on research papers written for a research audience (and where Wikipedia applications may not have been a direct focus). Even so, we believe that better integration of the Wikipedia community into current research as well as creating links or opportunities for feedback between editors and researchers would be mutually beneficial for both parties. We note that there have been previous efforts to engage the Wikipedia community more directly in the development of models such as through empowering editors to build datasets for model creation \cite{kuo2024wikibench}. We are far from the first to recognise the value that such interaction could bring for both parties (see for example, \cite{jemielniak2016bridging}). Nonetheless, we believe that better integration of editors into the design and development process could help offer insight into the most significant research challenges within Wikipedia, as well as in preventing duplication of effort or development of tools that the community may reject. 

\section{Future Work}

Based on the sampled literature, we note a number of research directions that future work could follow to build on and improve automatic generation of Wikipedia content. We identify three potential focuses: multilingual content, the interaction between content generation techniques and opportunities for evaluation.

\subsection{Multilingual Content}

While English is by far the largest language within Wikipedia, the service remains an inherently multilingual resource. Despite this, only 12 of the sampled studies applied and/or evaluated their methods to Wikipedia content beyond English and only 3 studies used more than two languages. Few language families were represented across the studies and generally, studies focused on the most widely spoken world languages although we note some exceptions to this such as Esperanto\footnote{An artificially constructed language estimated to have up to 2 million speakers} \cite{kaffee2018learning} and Nepali \cite{pariyar2014inconsistency}. We believe that this risks increasing the divide between high-resourced and under-resourced or underserved languages for which there is a smaller community of editors and/or users. While information in Wikipedia can be shared or transmitted across languages \cite{roy2022information}, growth and behaviour patterns in language editions can be very diverse and are not related to or constrained by issues of geography or linguistic characteristics \cite{ban2017robust}. Adding content to one language, then, does not guarantee that it will reach other languages even if they would outwardly appear to be similar. 

We also note that source materials used within the sampled studies often prevent the approaches from being applied to other languages. Most of the studies focused only on English and used only English articles and datasets for article generation. While Wikipedia features over 320 language editions at the time of writing, English is by far the largest. Given the differences in number, length and sometimes quality between English and other Wikipedia instances \cite{ban2017robust}, it is unlikely models could be generalised to other languages without significant adjustments and changes in model performance. We also question what this would mean for referencing and verifiability. While an LLM, for example, might be able to generate content in as many languages as -- or more languages than -- models trained from Wikipedia, it is unclear how or from where they would be able to source reference content. 

Wikipedia also features a number of lower-resourced languages such as African languages where the quantity and at times, quality of content greatly differs. Editors for these lower-resourced languages struggle to produce articles in part due to a lack of suitable source content in their languages \cite{nigatu2024low}. Such language communities would clearly benefit greatly from tools to assist in the creation of articles. However, we recognise that the lack of source materials may be a significant barrier to the development of such tools. We nonetheless highlight this as an interesting and important area for future work. We also believe that Wikidata poses an important opportunity for multilingual content models due to its language independent nature, the verifiability of Wikidata content and its knowledge-graph based structured data format.

\subsection{Interaction of Models}

Despite the wide variety of models and approaches identified through our analysis, even among the models intended to generate full-length articles from scratch, we found no models able to produce every form of content at once. To produce high quality articles that include all of the sections detailed in figure \ref{fig:article-outline}, multiple automated agents would be required to collaborate on the production of the article. Yet, unsurprisingly, each of the studies evaluated a single proposed approach in isolation rather than in conjunction with other automated agents. Some such as \citet{banerjee2015filling} did upload their outputs to Wikipedia and it is plausible that at least some of the subsequent edits made to this content were partially made by automated agents, but these studies do not describe these edits in sufficient data to evaluate the likelihood and nature of such interactions.

Interactions between bots have been shown to largely be collaborative and beneficial for the platform as a whole \cite{geiger2017operationalizing}. Even so, there remain tensions between some bots, particularly for bot developers who sometimes report unexpected interactions which require disabling or modifying bot behaviour \cite{geiger2017operationalizing}. Even where conflicts arise, these can lead to increased attention from -- and collaboration between -- editors to find the most appropriate content for an article \cite{geiger2023wikipedia}. Regardless, we believe that understanding how bots will interact and proposing methods to facilitate, manage and audit this interaction is a key area for future work and an issue that must be addressed if automated agents are to effectively collaborate on creating and improving content. 

\subsection{Evaluation}

We believe that the evaluation of -- and evaluation methods for -- automatically-generated content represent a key direction for future work. The heavy focus within the studies on automated content generation and the fluency of text fails to account for Wikipedia's content rules and the many ways in which perfectly fluent text could be entirely unsuitable for Wikipedia. 

Notably, the Wikipedia community has clearly shown a resistance to the introduction of algorithmic tools and methods which do not leave scope for the engagement of editors and users \cite{smith2020keeping}. Additionally, the failure to engage the community risks that content may fall foul of Wikipedia's guidelines -- at least as interpreted by the community -- and be reverted or removed. For example, \citet{duan2023wikipedia} give the example of creating a biography for Japanese jazz pianist Yuriko Nakamura who has a Wikipedia page with limited information. Any such content would need to comply with Wikipedia's ``Biographies of Living Persons" policy\footnote{\url{https://en.wikipedia.org/wiki/Wikipedia:Biographies_of_living_persons}}, a complex and detailed policy which among other factors details the types of source which are and are not suitable. While the approach suggested by the authors appears to be effective and to have the capacity to play an important role in filling gendered content gaps in Wikipedia, it appears that it may not fully align with the policy as described at the time of writing due to a potential lack of external references.

As well as researching how the community perceives such content, however, we also believe there is scope for diversification of automated content generation methods, as well as opportunities to develop new methods which better account for content suitability. In their review, \citet{moas2023automatic} identified a wide range of machine learning techniques which can be applied to identify the quality of an article. As a starting point, we believe these methods should be applied to automatically evaluate content in contexts where user-engagement is not possible or practicable, as well as being developed further to allow for better understanding of what makes an article high quality. 

\section{Limitations}

We are aware of several limitations inherent in our approach. Firstly, while we tried to be broad with our choice of search terms, we cannot exclude the possibility of other, relevant papers which were excluded from our analysis due to differences in terminology. More broadly, our choice of methodology and reliance on a systematic literature review by necessity excludes automated content generation approaches which have not or cannot be published, most notably any active, retired or proposed bots for Wikipedia. We caution that our analysis is not exhaustive and while there have been past analyses of the roles and activities adopted by Wikipedia bots, exploration of how bots author content is an interesting area for further analysis. Our decision to focus solely on studies produced in English is also likely to have had a substantial impact on our results and analysis as we recognise that monolingual models produced in languages other than English may have been published in non-English speaking venues. This decision was made out of necessity with English studies selected as a compromise due to the difficulty with accessing and fairly evaluating publications outside of the languages spoken by the author team. This is not, however, to suggest that studies and methods in languages beyond English are not important and we recognise this as a weakness in our approach which could be addressed in future work.

Finally, our analysis, by necessity, relies largely on the sources we were able to identify and sample at the time that our review took place. This means that while there may be further studies planned or under review using novel methods such as LLMs, we were unable to include such studies. Moreover, the topics and issues raised are focused on the literature identified through our review process. It is therefore possible that we touch on topics addressed in wider literature but which are not at the intersection between Wikipedia and automated generation. 

\section{Conclusion}

In this paper, we performed a systematic review of the automated generation of Wikipedia content through an in-depth analysis of 51 papers drawn from four thousand. Our findings suggest that all content present within the main namespace of Wikipedia can be generated by automated agents, but this content is rarely generated simultaneously. Although these models predominantly rely on existing Wikipedia content, the success of this process is also highly dependent on the availability of suitable external references and on the information available either through Web searches or Wikidata. To date, the Wikipedia community has played a relatively limited role in the evaluation of these models, which has predominantly relied on quantitative metrics or more rarely human evaluation by non-experts.

One area which our analysis was unable to explore was the role that generative AI tools such as ChatGPT may have in the generation and editing of Wikipedia content. Additionally, there exist a number of tools and models already in use within Wikipedia that we were by necessity unable to explore. This is an area for further research and likely to be an area where approaches evolve over time. Nevertheless, we suggest that there are clear opportunities for machine learning and artificial intelligence agents to contribute to Wikipedia. However, the success of these contributions will hinge on the answers to more complex questions such as what role human editors should play and how to retain the collaborative, open approach that Wikipedia was founded on as automation increases.

\begin{acks}
This work was supported by the Engineering and Physical Sciences Research Council [grant number EP/Y009800/1], through funding from Responsible AI UK (KP0011) and by the European Union’s Horizon Europe research and innovation programme under grant agreement number 101058677.
\end{acks}

\bibliographystyle{ACM-Reference-Format}
\bibliography{references}

\end{document}